# Location and allocation problem of high-speed train maintenance bases

Boliang Lin[a,*], Xiang Li[a], Yuxue Gu[a], Dishen Lu[b]

[a] School of Traffic and Transportation, Beijing Jiaotong University, Beijing 100044, China

[b] School of Computer and Information Technology, Beijing Jiaotong University, Beijing 100044, China

**Abstract:** Maintenance bases are crucial for the safe and stable operation of high-speed trains, necessitating significant financial investment for their construction and operation. Planning the location and task allocation of these bases in the vast high-speed railway network is a complex combinatorial optimization problem. This paper explored the strategic planning of identifying optimal locations for maintenance bases, introducing a bi-level programming model. The upper-level objective was to minimize the annualized total cost, including investment for new or expanding bases and total maintenance costs, while the lower-level focused on dispatching high-speed trains to the most suitable base for maintenance tasks, thereby reducing maintenance operation dispatch costs under various investment scenarios. A case study of the Northwest China high-speed rail network demonstrated the application of this model, and included the sensitivity analysis reflecting maintenance policy reforms. The results showed that establishing a new base in Hami and expanding Xi'an base could minimize the total annualized cost during the planning period, amounting to a total of 2,278.15 million RMB. This paper offers an optimization method for selecting maintenance base locations that ensures reliability and efficiency in maintenance work as the number of trains increases in the future.

## 1 Introduction

The decision regarding the placement of maintenance facilities plays a crucial role in the transportation system and strategic planning within contemporary industry. This decision necessitates a comprehensive assessment of various factors, including the superiority of the geographical location, the allocation of market demand, and the anticipated development trends. It represents a significant investment and profoundly influences the operational efficiency, stability, and safety of network operations. Therefore, the process of determining the optimal location for maintenance facilities is often a system engineering task involving complex combinatorial optimization.

In the context of high technology and high investment, such as the high-speed railway of China, the issue of location and scale for maintenance bases becomes particularly crucial. China State Railway Group Co., Ltd. (CSRG) adopts a typical preventive maintenance (PM) mode in the high-speed railway system (Wang [1]). This approach necessitates that high-speed electric multiple unit (EMU) trains return to a designated maintenance site for appropriate servicing before they reach specific time or mileage thresholds.

Both the depot and the maintenance base are facilities that undertake the maintenance and testing of EMU trains, and are essential for ensuring that all trains remain in good and safe operating condition. And they are differentiated based on the levels of maintenance they provide. China EMU maintenance plan is structured similarly to the common strategy of dividing locomotive and rolling stock maintenance into low-level and high-level categories, as outlined by Andres et al. [2]. This plan is segmented into five distinct levels. Level I and II maintenance, referred to as operational maintenance, are carried out in the depot, featuring high frequency and short duration. Level III, IV, and V maintenance, generally carried out by the maintenance base or manufacturer, are called high-level maintenance, which have lower frequency and longer duration. Using the CRH380B(L) series train as an example, Figure 1 illustrates the high-level maintenance cycle.

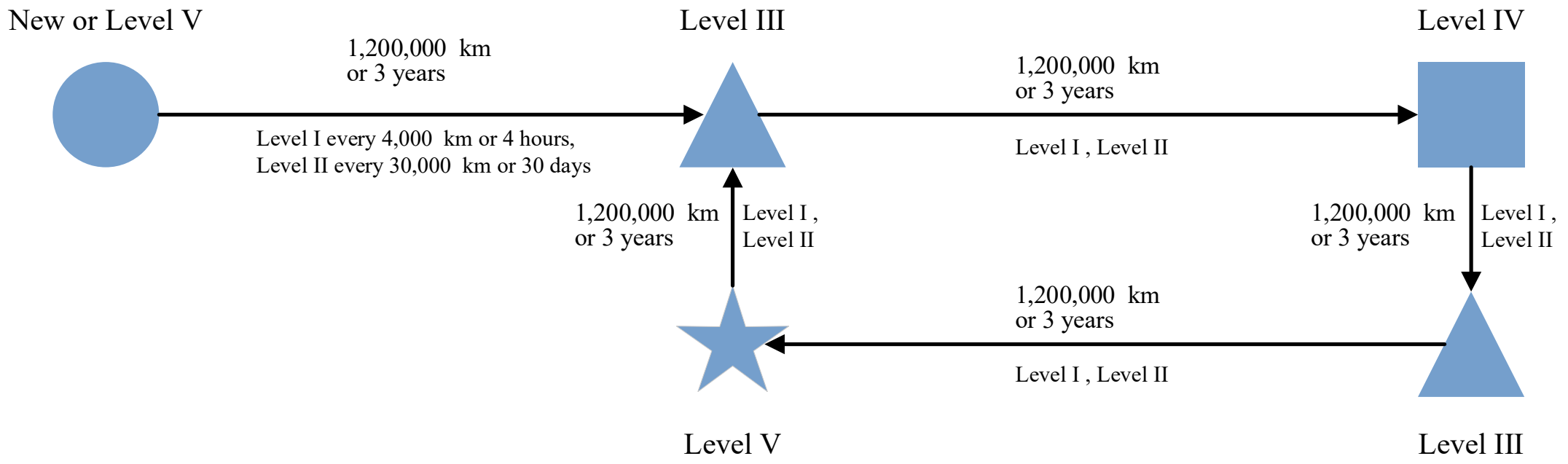

**Figure 1 High-level maintenance cycle for CRH380B(L) series train**

EMU high-level maintenance plans in China involves a systematic and hierarchical approach, aligning with three primary planning types: strategic, tactical, and operational. The process begins with annual strategic planning, rooted in business objectives and major influencing factors, to craft outline plans. Then, based on the outline plans and actual conditions, monthly high-level maintenance dispatch plans are developed, determining the specific maintenance dates. Finally, Operational plans are developed, continuously adapting to these specified dates for efficient train utilization and coordinating maintenance workshop activities. The frequency of operational maintenance is high, with intervals ranging from several hours to a few days, whereas the interval for high-level maintenance is longer, usually spanning several years. And maintenance bases, exemplified by Wuhan Maintenance Base, are notable for their extensive scale and significant investment. Wuhan base, initiated in 2007, spans 1.4 million square meters, with a construction area of 180,000 square meters, and represents a strategic investment of several hundred million USD. This highlights the strategic nature of planning in this context, focusing on long-term infrastructure and resource allocation to support the sustained operation and maintenance of EMU trains.

The concept of high-level maintenance, in contrast to operational maintenance, demands more rigorous and reliable standards for railway yards, necessitating a more scientific approach in the strategic placement of maintenance bases. With the growing number of high-speed trains and their increasing mileage, there is a rising need for high-level maintenance. Current maintenance bases might fall short in addressing the future requirements of a stable and efficient high-speed rail network. Thus, each potential investment plan and the choice of location for these bases could lead to different benefits and challenges. This necessitates a thorough evaluation of various factors such as the types of EMU trains, the benefits of operation and maintenance, and budgetary limitations.

This paper adopts a combinatorial optimization modeling approach to investigate methods for the maintenance base location-allocation problem (MBLAP). The remainder of this paper is as follows. Section 2 provides a detailed literature review of current relevant studies. Section 3 introduces the basic types of maintenance bases and analyzes the factors affecting the high-level maintenance demand and capability. Section 4 constructs a bi-level planning model, with the upper-level aiming to minimize the capital investment costs for new or expanded maintenance bases, and the lower-level minimizing the operational costs of high-speed trains for different investments, including the dispatch and maintenance operation costs generated by allocating process, and establishes an integer linear programming model. Section 5 presents a case study based on forecast data of the high-speed rail network in Northwest China, evaluating different scenarios under factors change. Finally, the paper concludes with conclusions and future research prospects.

## 2 Literature Review

Facility location problem (FLP) is one of the core issues in operations research. A common category in the study of FLP includes the uncapacitated facility location problem (UFLP), with relevant studies such as Hernandez et al. [3]. Based on practical needs, FLP has also been expanded to include variations such as the capacitated facility location problem (CFLP) (Sankaran [4]; Miao and Yuan [5]) and the multi-product facility location problem (MPFLP) (Nezhad et al. [6]). FLP focuses on determining the optimal geographical location of facilities to minimize costs or maximize service efficiency, while location-allocation problem (LAP) further considers how to efficiently allocate demand points to these facilities. LAP research covers a variety of complex application scenarios, with early studies including Cooper et al. [7] who proposed exact extremal equations and a heuristic method for solving LAP. Cebecauer and Buzna [8] also proposed a general adaptive aggregation framework for facility LAP issues. For maintenance facilities, Wu et al. [9] studied the facility allocation and maintenance strategy in cellular network planning based on heuristic algorithm and simulation. Holik [10] conducted a literature review from the perspective of departments of transportation. He pointed out that maintenance facilities have a long service life and their positioning needs to meet current and future needs.

In the railway transportation system, the reasonable arrangement of maintenance facilities is crucial to ensure the safety, stability, and economy of the system. The decision problem of maintenance facility selection and modification for locomotives and rolling stock can be considered as a precursor to MBLAP. Many scholars have studied this issue, with some focusing on the tactical level, which requires combining facility location planning with the locomotive routing problem. For example, Canca and Barrena [11] established an integrated locomotive and depot location model in the rapid rail system and employed Gurobi for its solution. Zomer et al. [12] constructed a mixed integer linear programming (MILP) model for the maintenance location choice problem, aiming to reduce capacity issues during the night. Castillo et al. [13] developed an optimal maintenance and operations scheduling model that ensures fleet availability to meet operational demands, while considering resource and time constraints at maintenance depots. Similar research exists in the aviation field, as aircraft also require regular maintenance. Gopalan [14] considers the task of locating the minimum number of maintenance bases necessary to ensure that all aircraft maintenance activities are carried out within a specified cycle. Stern and Saltzman [15] integrated the Aircraft maintenance routing problem (AMRP) with the aircraft maintenance facility location problem, creating an optimization model based on flight schedules.

This research at the tactical level usually focuses on route cycling times over several days. In context of China, this relates to operational maintenance tasks and depot location issues aligned with EMU train routing plans. But high-level maintenance tasks carried out by maintenance bases belong to the strategic level. At this level, Xie et al. [16] examined a multitype CFLP for optimizing North American locomotive maintenance workshop locations and capacities Tönissen and Arts [17] shifted from modeling daily individual train maintenance routing to an aggregated approach, using expected annual costs for routing trains to maintenance facilities. They developed a model for locomotive vehicle maintenance locations considering both present and future scenarios and created an efficient column-and-constraint generation algorithm for solving it. Later, Tönissen et al. [18] addressed the challenge of siting maintenance facilities in a railway network, considering uncertain line planning and other variables.

In terms of location and allocation, some studies are based on a phased strategy. For example, Harris et al. [19] proposed a two-stage planning method considering cost-effectiveness and distribution environment impact. Rostami et al. [20] chose to determine facility locations in the first phase and carry out allocation in the second phase. Li et al. [21] studied a capacitated facility location and protection problem using a two-stage stochastic bilevel model, addressing decision-dependent uncertainty in post-interdiction capacity. For the problem of railway multi-classification yard locations, Lin et al. [22] developed a bi-level programming model where the upper-level identifies the best construction investment plan for candidate nodes, based on yard size and capacity determined at the lower level, and utilizes a simulated annealing algorithm for computation. Tönissen and Arts [17] mentioned earlier also adopted a two-stage research approach, where the first stage determines the location and size of the maintenance facilities, while the second stage addresses an NP-hard maintenance location routing problem.

Beyond choosing locations for maintenance facilities, formulating EMU maintenance plans is another critical research area. Idris et al. [23] conducted an extensive study on the maintenance costs of rolling stock. While there is substantial research on routine lower-level maintenance (Giacco et al. [24]; Mira et al. [25]; Lin and Zhao [26]), high-level maintenance has received less attention due to its infrequent and longer cycles. Sriskandarajah et al. [27] examined maintenance overhaul scheduling for metro trains using genetic algorithms; however, their proposed reconfiguration plans faced practical implementation challenges. Wu et al. [28] developed a time-state network model to minimize high-level maintenance costs, considering passenger demand, workshop capacity, and maintenance regulations. Subsequently, Wu et al. [29] proposed a model for optimizing high-level maintenance planning to maximize usage efficiency, reducing unused mileage, and introduced a related particle swarm algorithm. Lin et al. [30] innovatively used a state function to indicate the EMU train maintenance status and applied simulated annealing for problem-solving. The latest research by Lin et al. [31] presents a 0-1 integer programming model for scheduling high-level maintenance, incorporating an algorithm that accounts for variable maintenance dates and daily mileages.

In summary, these literatures not only cover the location issues of maintenance bases in various types of networks and various methods and algorithms, but also include research on high-level maintenance of locomotive vehicles. Most of them choose to establish integer programming models to solve the location problems. However, there is currently a scarcity of literature specifically discussing the high-level maintenance facility location and the maintenance task allocation of EMU trains. And there is also a lack of consideration for the constraints of the variable capacities brought about by different construction plans, as well as the challenges in capacity sharing and consolidation in multi-product multi-facility problem.

The contributions of this paper are as follows:

1. We propose a bi-level programming mathematical model for selecting strategic investment plans for the new construction or expansion of maintenance base facilities, as well as the dispatch of EMU trains for maintenance from depots to maintenance bases.

2. The base location selection process considers multiple types of EMU trains as well as different needs for levels III, IV, and V high-level maintenance.

3. A no-investment plan is incorporated into the model as a special scheme, and a capital recovery coefficient is introduced to annualize capital investments.

4. The model and algorithm performance are verified on a real-world example of the high-speed railway network in northwest China, providing suggestions for the planning of maintenance base locations for CSRG.

## 3 Problem Description

In China, depots are responsible for the daily storage and frequent operational maintenance of EMU trains, and are usually located near and connected to major city hub stations. Each train is assigned to a depot, and when they reach the threshold for high-level maintenance, they need to travel from their assigned depot to the maintenance base for high-level maintenance operations, and then return to their assigned depot after the maintenance is completed.

It is obviously that the farther the average distance between the train assigned points and maintenance bases, the emptier travel mileage is generated (of course, for trains dispatched over long distances, it is also possible to undertake some transport tasks using the non-productive travel path when conditions are mature, but this is currently not common). Bases need to be selected based on limited budgetary and geographical location options, often resulting in a limited number, forming a many-to-one relationship with the depots. At this point, it is necessary to consider the high-level maintenance demand of each depot, avoiding as much as possible being far from depots with many assigned trains to prevent excessive non-productive travel distances for long-distance maintenance. Besides, the maintenance capacity of each base should be well-coordinated with others and aligned as closely as possible with the quantity and types of EMU trains assigned to the nearby depots.

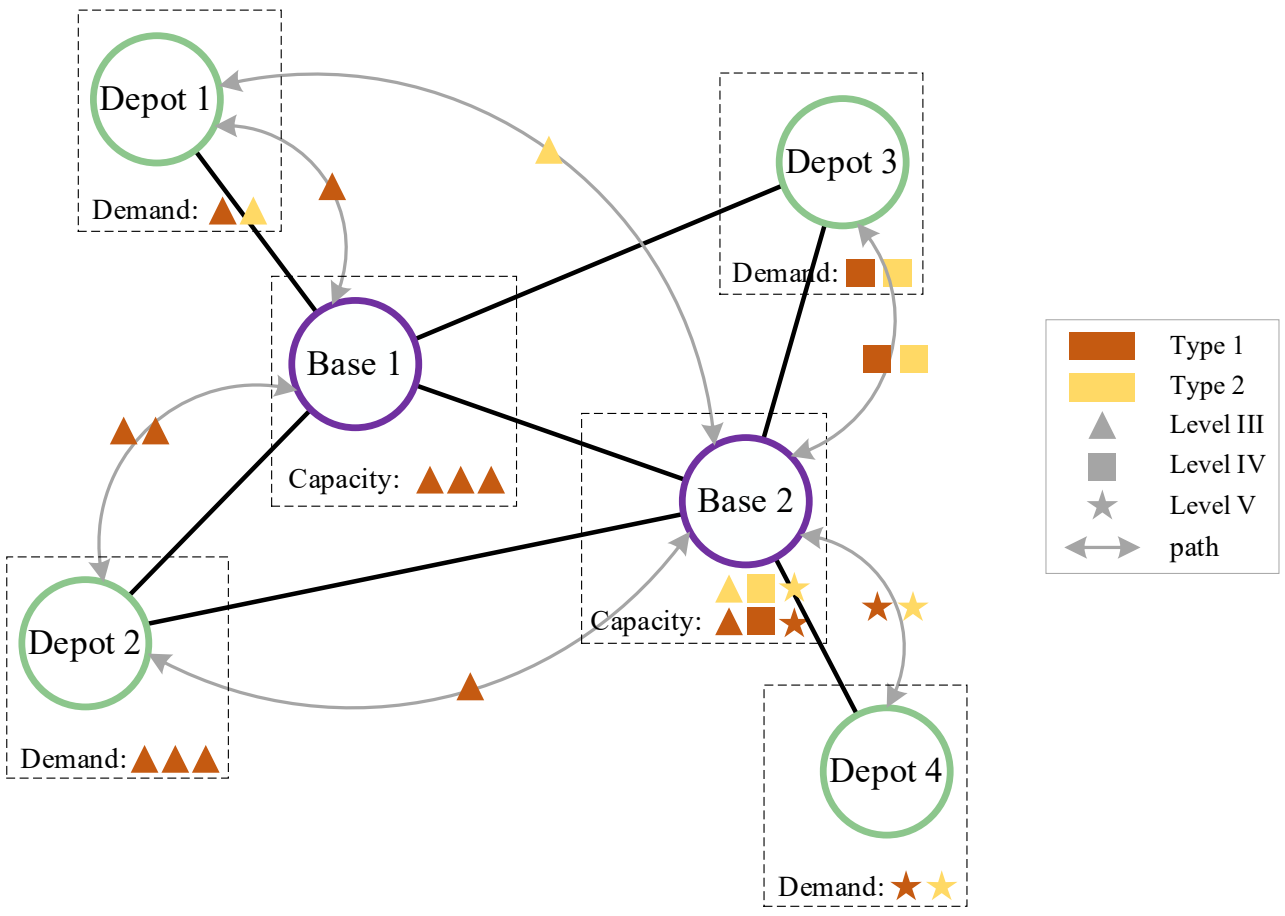

**Figure 2 An example of high-speed rail network and maintenance allocation**

Figure 2 showcases a high-speed rail network example featuring 2 maintenance bases and 4 Depots. Once a maintenance plan for a specified duration is established, the directional arcs illustrate the dispatch and return relationships of EMU trains for maintenance. Notably, due to the lack of required maintenance capabilities at Base 1, Type 2 trains assigned to Depot 1 must be sent to the farther Base 2 for maintenance. Similarly, as the level III capacity of Base 1 for Type 1 trains is already at its limit, one of the three Type 1 trains needing maintenance from Depot 2 also has to travel to the more distant Base 2.

In this example, the capacities of both bases are saturated. If new trains are assigned to this region, and if it is decided to maintain them within the region instead of sending them back to the manufacturer, this creates a dual problem about existing maintenance resources. The first problem entails determining the location, scale, and type of resource expansion, while the second revolves around refining the dispatch strategy for trains to achieve an efficient location-allocation alignment.

## 3.1 Capacity of maintenance bases

To determine the investment and construction capabilities for a maintenance base, it is necessary to consider the specific operational processes of high-level maintenance and the internal spatial structure of the base. High-level maintenance typically involves several stages, including pre-inspection, lifting, bogie repairing, cleaning, lowering, testing and final preparation. For level III maintenance, the standard procedure is to lift the entire EMU train, but levels IV and V maintenance processes are more complex, requiring the disassembly of the train into individual cars. These levels involve more detailed and thorough maintenance than level III, including tasks such as repainting.

Figure 3 provides a conceptual overview of typical high-level maintenance facilities and processes, which may vary depending on the maintenance level. Facilities for level III maintenance are relatively simple, including not only the high-level maintenance shop but also the general-purpose static testing shop and bogie maintenance shop, as illustrated in Figure 3 (a). Facilities required for levels IV and V maintenance are more comprehensive and are designed for compatibility between these two levels. Building on the level III facilities, necessary additional facilities such as the car body shop, the component shop, and the painting shop are included, as depicted in Figure 3 (b).

Therefore, from the perspective of investment efficiency, maintenance bases usually adopt a phased construction strategy, i.e., initially only constructing level III maintenance facilities. As the quantity and scale of EMU trains increase and their maintenance requirements evolve, these bases undergo timely expansions and renovations to boost their capacity.

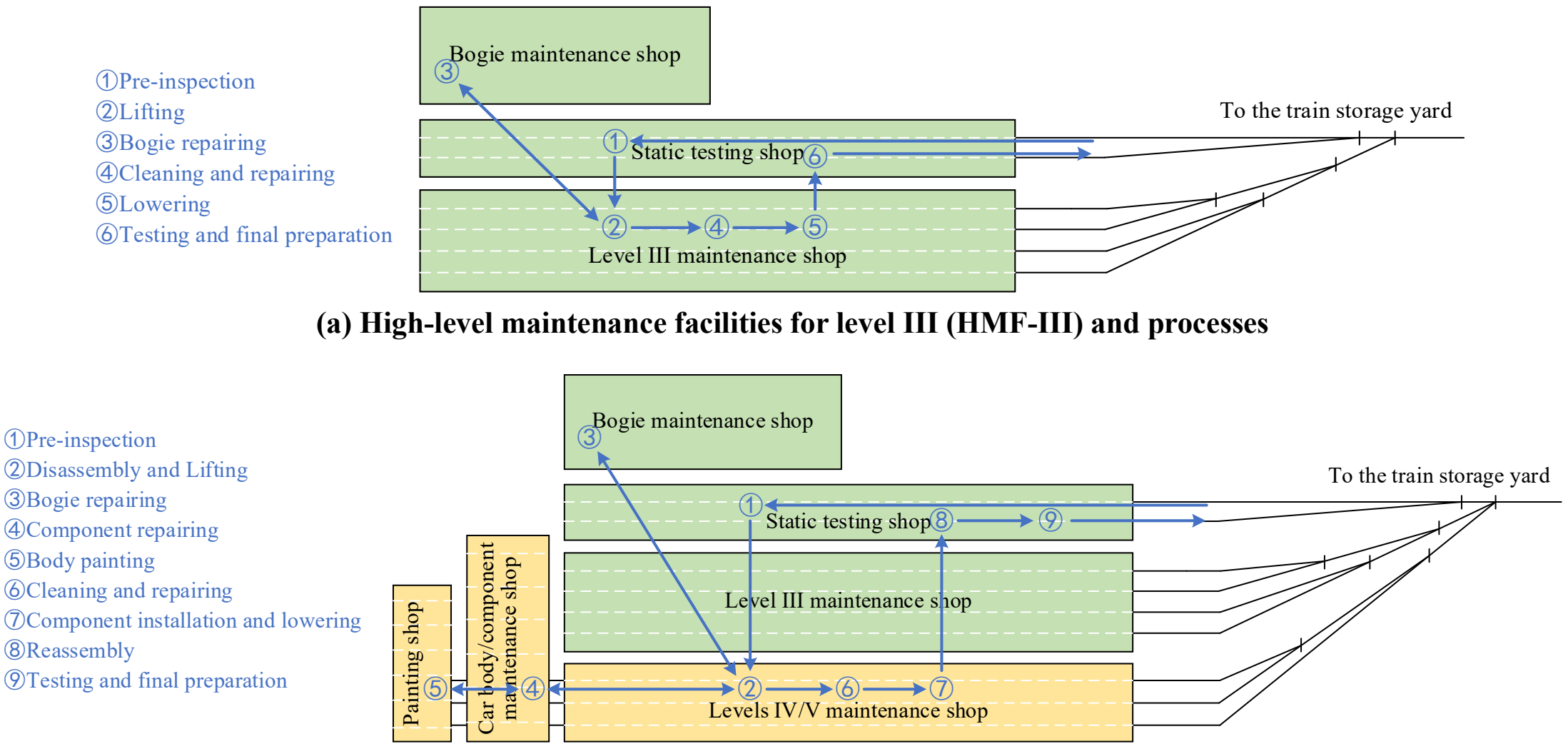

**(a) High-level maintenance facilities for level III (HMF-III) and processes**

**(b) High-level maintenance facilities for levels III&IV&V (HMF-III&IV&V) and processes**

**Figure 3 Typical high-level maintenance facility**

A maintenance line designed to service an 8-car high-speed EMU train (1 standard trainset) usually includes one maintenance position. In contrast, a line for a longer high-speed EMU train (2 standard trainsets), includes two maintenance positions. In other words, one position can undertake the maintenance tasks for one standard trainset.

The overall scale of high-level maintenance primarily considers the position number of maintenance lines in both the high-level maintenance shop and the static testing shop, as well as the service capacity of the bogie shop. For ease of calculation and reflecting actual usage, all mentioned maintenance activities could broadly be categorized into two types: lifting in the high-level maintenance shop and testing in the static testing shop, with a time allocation ratio of approximately 2:1.

In line with this, all construction investment plans proposed in this paper suggest a 2:1 ratio for the number of positions allocated to lifting and testing. This approach aims to prevent operational conflicts between these two essential tasks. The maintenance duration for bogies varies across different train types, making the quantification of production scale a complex task. Hence, this study bases its capacity determination on the aggregate number of positions across the maintenance lines without the consideration of the bogie. For instance, if each maintenance line in Figure 3 (a) has two positions, with four maintenance lines in the level III maintenance shop and two in the static testing shop, then the facility would have a total of six maintenance lines (equating to 12 positions) available for providing maintenance capacity.

In considering the cost variations of maintenance bases, it is essential to account for the impact of capacity settings from the perspective of maintenance operations and shop layout. Additionally, from a business management viewpoint, under identical conditions, a higher maintenance quote for a base tends to result in a decrease in the number of orders received. Factors such as local land prices, topography, the level of local economic development, and labor costs can lead to significant differences in construction and operational expenses. This is particularly true for bases of the same capacity located in different areas, substantially influencing the optimization process.

### 3.2 Demand of EMU high-level maintenance

The demand for high-level maintenance of trainsets is positively correlated with the fleet size of EMU trains, and it also depends on factors such as different EMU types and the maintenance content and cycle of different levels. As illustrated in Figure 1, to avoid unnecessary over-maintenance, the schedules for various maintenance levels often have a multiplicative relationship. Clearly, due to variations in the manufacturing dates and usage patterns of trainsets, there are inevitable fluctuations in the need for high-level maintenance. Planning maintenance activities over different periods is a complex task (Budai [32]). Nonetheless, when analyzing enough trains over a time frame that surpasses their lifespans, the annual average demand for high-level maintenance tends to be stable, despite some fluctuations.

Moreover, as the high-speed railways operation stabilizes in China, and significant reassignments of numerous trainsets become infrequent, and CSRG usually set a maximum limit for the number of trainsets undergoing maintenance each month. Therefore, by incorporating an imbalance coefficient, it becomes feasible to approximate the annual high-level maintenance needs for trainsets at each depot.

## 4 Bi-level programming model for MBLAP

In the realm of maintenance base construction, there are primarily two strategies: upgrading and expanding existing bases through technological enhancements, or establishing entirely new bases at different locations. These strategies can be implemented concurrently in various locations. The optimization of this process involves a dual-variable approach: firstly, deciding whether to construct a base, which includes expansions via technological upgrades; and secondly, allocating the planned maintenance EMU trains from each depot to the appropriate bases. The primary objective of this model is to minimize the overall costs. This paper, therefore, presents a two-stage mathematical model that encapsulates these considerations.

### 4.1 Symbol definition

The definitions of sets, parameters, and other relevant symbols involved in the model are shown in Table 1.

**Table 1 Definitions of symbols**

| Symbol | Definition |
|---|---|
| $S^{\text{Base}}$ | Set of bases (including existing bases and potential alternative bases) |
| $S_j^{\text{Project}}$ | Set of construction projects (new construction or expansion) available for base $j$ |
| $S^{\text{Depot}}$ | Set of Depots |
| $S^{\text{Type}}$ | Set of EMU types |
| $S^{\text{Level}}$ | Set of high-level maintenance grades for EMU trains, $S^{\text{Level}} = \{3,4,5\}$ |
| $T_j^p$ | Payback period for investment when base $j$ chooses plan $p$ |
| $I_j^p$ | Construction investment required for base $j$ when selecting plan $p$ |
| $B$ | Total investment budget for construction of bases during the planning period |
| $\gamma$ | Annual interest rate on funds |
| $N_i^{eg}$ | Annual number of standard trainsets for type $e$ EMU trains assigned to depot $i$ that require level $g$ maintenance |
| $\mathbb{C}_j^{eg}$ | Initial maintenance capacity of base $j$ at the start of the planning period, which is the number of standard trainsets for type $e$ EMU trains that can be serviced annually for level $g$ maintenance after conversion |
| $\Delta\mathbb{C}_{jp}^{eg}$ | Increase in level $g$ maintenance capacity for type $e$ EMU trains at base $j$ after investment renovation under construction plan $p$ |
| $\alpha$ | Cost incurred per kilometer of non-productive travel for a standard trainset |
| $L_{ij}$ | Travel distance from depot $i$ to base $j$ |
| $C_j^{eg}$ | Cost for base $j$ to undertake level $g$ maintenance for a standard trainset for type $e$ EMU trains |
| $\xi_j^{eg}$ | Matching coefficient, 1 if base $j$ can perform level $g$ maintenance for type $e$ EMU trains, otherwise 0 |
| $\mu_j^{eg}$ | Matching coefficient, 1 if base $j$ can initially perform level $g$ maintenance for type $e$ EMU trains at the start of the planning period, otherwise 0 |
| $\Delta\mu_{jp}^{eg}$ | Matching coefficient, 1 if alternative construction plan $p$ for base $j$ provides maintenance capacity for level $g$ of type $e$ EMU trains, otherwise 0 |
| $\lambda^{eg}$ | Coefficient for converting type $e$ EMU trains needing level $g$ maintenance into maintenance capacity |
| $\theta$ | Unbalance coefficient for dispatching repairs, a real number greater than 1. If each Depot gets balanced dispatching, then $\theta = 1$, otherwise adjust this value based on the maintenance experience of bases |
| $y_j^p$ | 0-1 decision variable, 1 if base $j$ adopts construction plan $p$, otherwise 0 |
| $f_{ij}^{eg}$ | Integer variable, number of standard trainsets for type $e$ EMU trains assigned to depot $i$ dispatched to base $j$ for level $g$ maintenance |

Different construction blueprints for maintenance bases come with varying investment costs. The process of sending EMU trains that require maintenance to various high-level maintenance facilities results in different distances for non-productive travel and associated costs. Strategically locating these maintenance bases, coupled with the efficient dispatch of EMU trains from depots to the most suitable facilities, can significantly reduce the overall comprehensive costs for the entire system.

4.2 Upper-level model for location and allocation

Based on the definition of the investment decision variables, for any base, the construction plan chosen has exclusivity, meaning there is an incompatibility constraint:

$$\sum_{p \in S_j^{\text{Project}}} y_j^p \le 1, \quad \forall j \in S^{\text{Base}} \tag{1}$$

In fact, not undertaking construction can also be considered a special type of investment plan, which can be marked as "0", with its corresponding investment cost $I_j^0 = 0$. This is represented by the decision variable $y_j^0$. Therefore, inequality constraint (1) can also be transformed into an equality constraint:

$$y_j^0 + \sum_{p \in S_j^{\text{Project}}} y_j^p = 1, \quad \forall j \in S^{\text{Base}} \tag{2}$$

Let the set of investment decision variables for base $j$ be denoted as $Y_j = (y_j^0, y_j^1, ..., y_j^{|S_j^{\text{Project}}|})$, then the decision vector for the investment and location plans of all bases within the entire regional network is represented as $Y = (Y_1, Y_2, ..., Y_j, ..., Y_{|S^{\text{Base}}|})$.

In theoretical terms, a maintenance base, as a fixed asset, will have a residual value of zero when its useful life is exhausted. For instance, Stern and Saltzman (2023) annualize the total construction cost of facilities by dividing the total construction cost by the expected facility life, but such an approach might overestimate the annualized cost. In practical operations, due to the updating and upgrading of facilities and equipment, a base built under normal operating conditions is likely to exist indefinitely. Therefore, the capital recovery factor (CRF) $\lambda_j^{\text{Investment}}$, denoted as lambda, is introduced to calculate the annualized investment in the base. CRF also known as the "investment recovery factor" or "capital recovery coefficient", refers to the ratio of the annual recovery amount to the invested amount under the condition of compound interest over a predetermined recovery period. Under a given interest rate $\gamma$ and term $T_j^p$, the amount of money that needs to be recovered annually to fully recover the original investment is defined according to the used interest rate and number of years, and can be defined as:

$$\lambda_j^{\text{Investment}} = \gamma\left(1+\gamma\right)^{T_j^p} \Big/ \left[\left(1+\gamma\right)^{T_j^p} - 1\right] \tag{3}$$

In the process of selecting and optimizing the investment plan for maintenance bases, the comparison of plans needs to consider not only the construction costs of each plan, but also by aiming to minimize the costs associated with the non-productive travel of EMU trains for maintenance purposes. As a result, the optimization goal of the upper-level planning model is to minimize the sum of the annualized construction cost of the maintenance bases and the annual total maintenance cost over the planning period. Thus, the upper-level planning model of MBLAP can be described as follows:

$$\min Z^{\text{Upper}} = \sum_{j \in S^{\text{Factory}}} \sum_{p \in S_j^{\text{Project}}} \lambda_j^{\text{Investment}} I_j^p y_j^p + Z^{\text{Lower}}(Y) \tag{4}$$

(MBLAP_U) s.t.

$$y_j^0 + \sum_{p \in S_j^{\text{Project}}} y_j^p = 1, \quad \forall j \in S^{\text{Base}} \tag{5}$$

$$\sum_{j \in S^{\text{Base}}} \sum_{p \in S_j^{\text{Project}}} I_j^p y_j^p \le B \tag{6}$$

$$y_j^0, y_j^p \in \{0,1\}, \quad \forall j \in S^{\text{Base}}, p \in S_j^{\text{Project}} \tag{7}$$

The first component of objective function reflects the annualized investment cost adjusted by CRF; the second term represents the annual total maintenance cost (including maintenance operation cost and dispatch) for trains during the planning period. Specifically, $Z^{\text{Lower}}(Y)$ is the annual total maintenance cost under the optimal allocation plan, given a set of investment decision variables $Y = (Y_1, Y_2, ..., Y_j, ..., Y_{|S^{\text{Base}}|})$, with the specific value obtained through the lower-level planning. Constraint (6) represents the investment budget constraint, meaning that the funds allocated for base construction during the planning period cannot exceed the total budget $B$. And constraint (7) pertains to the nature of the variables.

It is quite clear that, under the premise of not considering dispatching costs and maintenance capabilities, choosing not to undertake any investment results in the minimum value of $Z^{\text{Upper}}$, which is also the optimal solution for the upper-level planning, i.e., for $\forall j \in S^{\text{Base}}$, there is $y_j^0 = 1$, $I_j^0 = 0$. Nevertheless, if the objective of the upper-level planning is considered as the final goal, and there is a decision not to invest in enhancing maintenance capabilities, it may result in situations where the demand for high-level maintenance of EMU trains cannot be met in some cases.

This would lead to the need for queuing in high-level maintenance scheduling and cause extended maintenance times, or force the trains to be sent to more distant external areas for maintenance, thereby reducing the efficiency of train utilization.

4.3 Lower-level model for maintenance operation and dispatch costs

It is assumed that high-level maintenance tasks for EMU trains assigned to chosen depots are carried out within the maintenance bases in the study area. These tasks are not outsourced to facilities outside the region or back to the EMU manufacturers. Under a set of construction plan variables $Y$ already determined by the upper-level planning, dispatching EMU trains should incur lower execution costs. The lower-level planning model for dispatching trains to maintenance bases for dispatch can be described as follows:

$$\min Z^{\text{Lower}} = \alpha \sum_{i\in S^{\text{Depot}}} \sum_{j\in S^{\text{Base}}} L_{ij} \sum_{e\in S^{\text{Type}}} \sum_{g\in S^{\text{Level}}} f_{ij}^{eg} + \sum_{j\in S^{\text{Base}}} \sum_{g\in S^{\text{Level}}} \sum_{e\in S^{\text{Type}}} C_j^{eg} F_j^{eg} \tag{8}$$

(MBLAP_L) s.t.

$$\sum_{j\in S^{\text{Base}}} \xi_j^{eg} f_{ij}^{eg} = N_i^{eg}, \quad \forall i\in S^{\text{Depot}}, e\in S^{\text{Type}}, g\in S^{\text{Level}} \tag{9}$$

$$\xi_j^{eg} \ge \mu_{jp}^{eg} + \sum_{p\in S_j^{\text{Project}}} y_j^p \Delta\mu_{jp}^{eg} - \mu_{jp}^{eg} \sum_{p\in S_j^{\text{Project}}} y_j^p \Delta\mu_{jp}^{eg}, \quad \forall i\in S^{\text{Depot}}, j\in S^{\text{Base}}, e\in S^{\text{Type}}, g\in S^{\text{Level}} \tag{10}$$

$$\xi_j^{eg} \le \mu_{jp}^{eg} + \sum_{p\in S_j^{\text{Project}}} y_j^p \Delta\mu_{jp}^{eg}, \quad \forall i\in S^{\text{Depot}}, j\in S^{\text{Base}}, e\in S^{\text{Type}}, g\in S^{\text{Level}} \tag{11}$$

$$\theta \sum_{e\in E} \sum_{g\in G} \lambda^{eg}(E,G) F_j^{EG} \le \mathbb{C}_j^{EG} + \sum_{p\in S_j^{\text{Project}}} y_j^p \Delta\mathbb{C}_{jp}^{EG}, \quad \forall j\in S^{\text{Base}}, E\in S^{\text{Type}}, G\in S^{\text{Level}} \tag{12}$$

$$f_{ij}^{eg} \in \mathbb{N}, \quad \forall i\in S^{\text{Depot}}, j\in S^{\text{Base}}, e\in S^{\text{Type}}, g\in S^{\text{Level}} \tag{13}$$

In the lower-level planning model, the first term of the objective function represents the cost incurred by all trains traveling empty to the base for maintenance and returning to the depot. The second term is the annual high-level maintenance operation cost for trains, where the number of standard trainsets for type $e$ EMU trains undergoing level $g$ maintenance annually at maintenance base $j$, denoted as $F_j^{eg}$, can be calculated as follows:

$$F_j^{eg} = \sum_{i\in S^{\text{Depot}}} f_{ij}^{eg} \tag{14}$$

In the constraints, constraint (9) is a condition for flow balance, indicating that the total number of standard trainsets requiring maintenance generated by all depots is equal to the number of maintenance tasks performed at all bases; constraints (10) and (11) together stipulate that after completing construction plan $p$, base $j$ has the capability to undertake maintenance tasks of the corresponding type and level; constraint (12) limits the maintenance capacity of base $j$ after completing the new construction or renovation under plan $p$. In practical operations, due to limitations in maintenance resources and the need for “compatible maintenance”, a single maintenance line can usually handle several different types or levels. At this time, several types $e$ or levels $g$ can be referred to as $E$ and $G$ after being combined. In such cases, a conversion coefficient function, depending on the combination situation, needs to be considered when calculating the capacity. If no compatibility is involved, then constraint (12) can also be expressed as:

$$\theta \sum_{e\in S^{\text{Type}}} \sum_{g\in S^{\text{Level}}} F_j^{eg} \le \mathbb{C}_j^{eg} + \sum_{p\in S_j^{\text{Project}}} y_j^p \Delta\mathbb{C}_{jp}^{eg}, \quad \forall j\in S^{\text{Base}} \tag{15}$$

and constraint (13) specifies that the variables are non-negative integers.

4.4 Mixed integer programming model for MBLAP

In addition to the layered modeling approach, the above-mentioned upper-level and lower-level models can also be combined into an integrated model, as follows:

$$\begin{aligned} \min Z = & \sum_{j\in S^{\text{Base}}} \sum_{p\in S_j^{\text{Project}}} \lambda_j^{\text{Investment}} I_j^p y_j^p \\ & + \alpha \sum_{i\in S^{\text{Depot}}} \sum_{j\in S^{\text{Factory}}} L_{ij} \sum_{e\in S^{\text{Type}}} \sum_{g\in S^{\text{Level}}} m^e f_{ij}^{eg} \\ & + \sum_{j\in S^{\text{Base}}} \sum_{g\in S^{\text{Level}}} \sum_{e\in S^{\text{Type}}} C_j^{eg} F_j^{eg} \end{aligned} \tag{16}$$

（MBLAP） s.t.

（5）-（7），（9）-（13）

Model MBLAP involves 0-1 variables and integer variables, making it a mixed integer programming model in mathematical terms. Its objective function and constraints are a combination of those in MBLAP_U and MBLAP_L.

4.5 High-level maintenance demand and maintenance base capacity

Due to the high utilization rate of EMU trains in China, they generally reach the mileage threshold first. Therefore, $N_i^{eg}$ can be calculated based on mileage cycles as:

$$N_i^{eg} = \omega m^e \left[ H_i^e l^e / L^{eg} - \left(1-\delta_{g,5}\right) H_i^e l^e / L^{eg+1} \right], \quad \forall i \in S^{\text{Depot}}, e \in S^{\text{Type}}, g \in S^{\text{Level}} \tag{17}$$

where $m^e$ is the standardization factor for the type $e$ EMU, with 8-car EMU trains taken as 1, and long EMU trains taken as 2; $H_i^e$ is the number of type $e$ EMU trains assigned to depot $i$; $l^e$ is the annual mileage of type $e$ EMU trains, taken as 65,000 km per year in practice; $L^{eg}$ is the mileage cycle for level $g$ high-level maintenance of type $e$ EMU trains, the general ratio of $L^{e3}$ : $L^{e4}$ : $L^{e5}$ is 1:2:4 based on Figure 1 (in km); $\omega$ is the unbalance coefficient considering the fluctuation of annual high-level maintenance demands, simplified for calculation purposes, with demand fluctuation numerically represented by $\theta$, here taken as $\omega = 1$; $\delta_{i,j}$ is the Kronecker function, defined as follows:

$$\delta_{i,j} = \begin{cases} 1, & \text{if } i = j \\ 0, & \text{else} \end{cases} \tag{18}$$

The annual level $g$ maintenance capacity for type $e$ EMU trains at base $j$ not only considers the sum of lifting and testing positions but also needs to be calculated based on the maintenance duration, annual working days, and work unbalance coefficient. The calculation formula is:

$$\mathbb{C}_j^{eg} = P_j^{eg} D^{\text{Year}} / \alpha d^{eg}, \quad \forall j \in S^{\text{Base}}, e \in S^{\text{Type}}, g \in S^{\text{Level}} \tag{19}$$

where $P_j^{eg}$ is the number of positions at base $j$ capable of undertaking level $g$ maintenance for type $e$ EMU trains; $D^{\text{Year}}$ is the number of annual working days, set at 300d according to the working system of Chinese railway; $\alpha$ is the work unbalance coefficient considering reduced efficiency due to equipment failure, production organization, and abnormal parts supply, numerically represented by $\theta$, and here taken as $\alpha = 1$; $d^{eg}$ is the maintenance duration for level $g$ maintenance of type $e$ EMU trains, set at $d^{e3} = 30\text{d}$, $d^{e4} = 45\text{d}(\forall e \in S^{\text{Type}})$.

## 5 Case Study

5.1 Network and data

The high-speed railway network in Northwest China is managed by Urumqi, Qinghai-Tibet, Lanzhou, and Xi’an Railway Groups, effectively connecting important cities in the region. Considering the long payback period for infrastructure investments, and the need for forward planning, the case study analysis includes both existing lines and depots, as well as future under-construction and planned projects.

To facilitate resource sharing and cost-efficiency, maintenance bases are often upgraded from existing Depots. Accordingly, the selected high-speed railway network and facility distribution are shown in Figure 4. The region currently has seven Depots in Urumqi, Hami, Xining, Yinchuan, Lanzhou West, Baoji South, Xi'an North, and Xi'an East (with Hami, Baoji South, and Xi'an East as planned depots). However, the sole maintenance base in the area is Xi'an Maintenance Base, situated at Xi'an North Depot.

With the economic and social development in western China, and the continuous introduction of new high-speed rail lines, the number of EMU trains in the northwest region is expected to increase, and existing maintenance resources may not be sufficient to meet maintenance needs. Currently, only Xi'an base exists within this area, and EMU trains assigned to Urumqi depot, if sent to Xi'an base for maintenance, would have to travel thousands of kilometers empty (as shown by the distance arc in Figure 4). Optimizing the distribution of EMU maintenance resources in this region thus emerges as a highly pertinent issue.

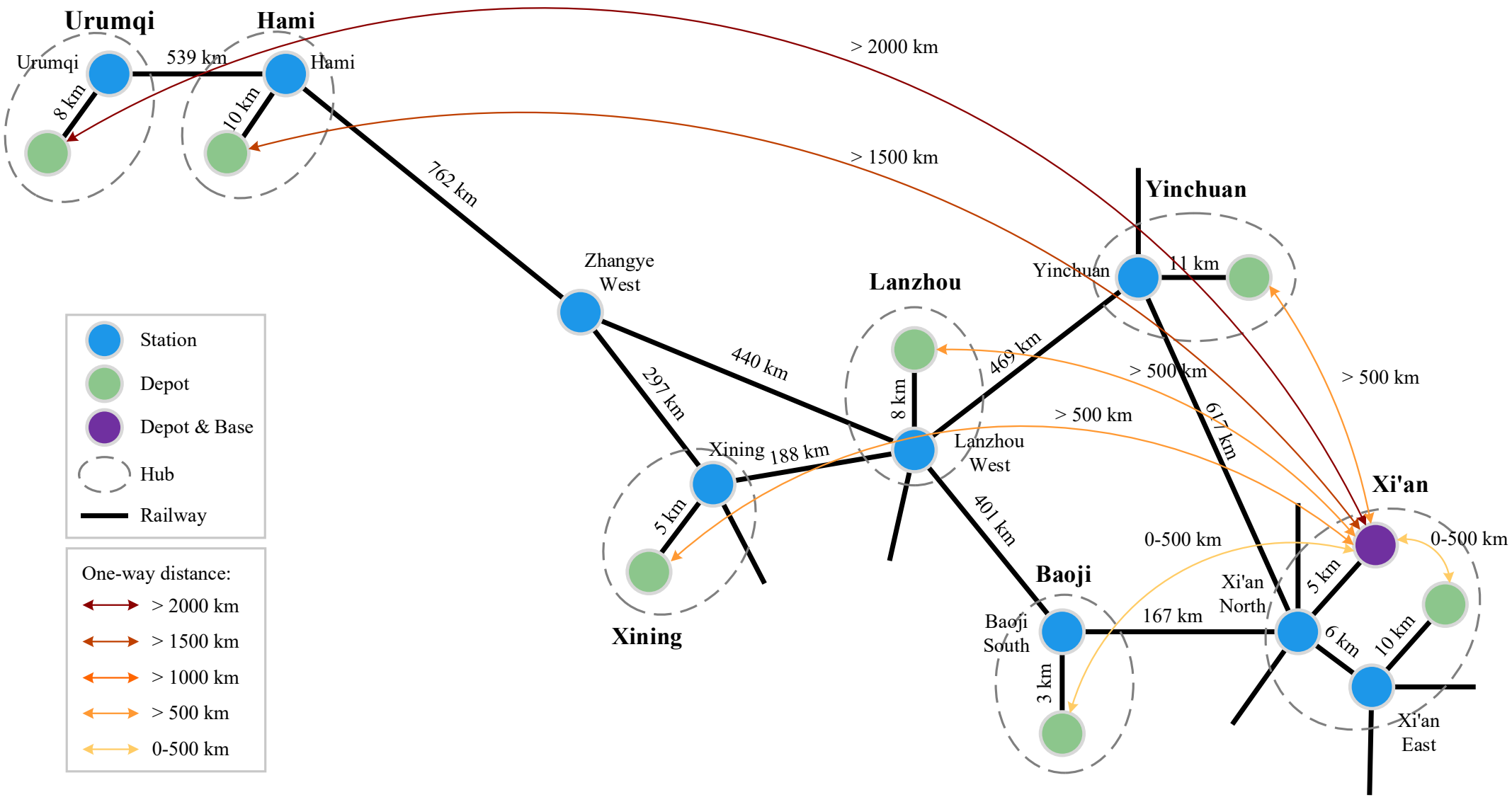

**Figure 4 Schematic diagram of the high-speed rail network**

Assuming the combined operational and opportunity cost for a single standard trainset traveling empty for 1 km is 1,500 RMB, And the investment payback period for all construction plans is 20 years, i.e., $T_j^p = 20$, with the basic interest rate is set at $\gamma$=6.5%, resulting in the calculation of CRF $\lambda_j^{\text{Investment}} \approx 9.1\%$. Based on experience, the dispatching unbalance coefficient is set at $\theta = 1.2$ Additionally, the total budget for base construction is assumed to be 5,000 million RMB, denoted as $B = 5,000$.

These depots are assigned EMU trains from both the CRH and CR series, encompassing 8 different types. Chinese manufacturers have largely ceased production of the CRH series, focusing primarily on the CR series. With the standardization of operation and maintenance standards within the same series, future maintenance facilities will only need to be divided into these two series. As shown in Table 2, by the end of the planning period, the total number of trains is expected to grow to 421 trains (451 standard trainsets). Since maintenance facilities usually have some capacity redundancy, this analysis overlooks the scenario where long EMU trains are "split" into two standard trainsets and sent to different bases, focusing instead on the total number of standard trainsets generated by each depot. For calculations, $L^{e3}$ is taken as 120,000 ($e = 1,2$), 145,000 ($e = 3,4,5,6$), and 165,000 ($e = 7,8$).

**Table 2 Number of EMU trains assigned to each depot**

| $i$ | Depot | $e$ | Type | $m^e$ | No. of trains | No. of CRH Std. sets | No. of CR Std. sets |
|---|---|---|---|---|---|---|---|
| 1 | Urumqi | 4 | CRH5G | 1 | 12 | 28 | 26 |
| | | 3 | CRH5E | 1 | 2 | | |
| | | 7 | CR300BF | 1 | 26 | | |
| | | 1 | CRH2G | 1 | 14 | | |
| 2 | Hami | 4 | CRH5G | 1 | 12 | 12 | 14 |
| | | 7 | CR300BF | 1 | 14 | | |
| 3 | Xining | 4 | CRH5G | 1 | 6 | 6 | 12 |
| | | 7 | CR300BF | 1 | 12 | | |
| 4 | Lanzhou West | 5 | CRH380B | 1 | 27 | 39 | 28 |
| | | 4 | CRH5G | 1 | 12 | | |
| | | 7 | CR300BF | 1 | 18 | | |
| | | 8 | CR400BF | 1 | 10 | | |
| 5 | Yinchuan | 5 | CRH380B | 1 | 13 | 19 | 23 |
| | | 4 | CRH5G | 1 | 6 | | |
| | | 7 | CR300BF | 1 | 17 | | |
| | | 8 | CR400BF | 1 | 6 | | |
| 6 | Baoji South | 7 | CR300BF | 1 | 10 | 0 | 16 |
| | | 8 | CR400BF | 1 | 6 | | |
| 7 | Xi'an North | 5 | CRH380B | 1 | 58 | 118 | 80 |
| | | 6 | CRH380BL | 2 | 10 | | |
| | | 2 | CRH380AL | 2 | 20 | | |
| | | 7 | CR300BF | 1 | 60 | | |
| | | 8 | CR400BF | 1 | 20 | | |
| 8 | Xi'an East | 7 | CR300BF | 1 | 10 | 0 | 30 |
| | | 8 | CR400BF | 1 | 20 | | |
| **Total** | | | | | **421** | **222** | **229** |

The service range of existing Xi'an base extends beyond the scope of this case study. Therefore, it is assumed that the maintenance facilities available for this case study at Xi'an base are: level III CRH, and level IV&V CRH maintenance lines, each with an availability of 1.5 lines (3 positions). This paper stipulates that each railway hub can construct a maximum of one base, with the potential locations for the alternative maintenance bases being the Depots within each hub (excluding Xi'an East). The facilities for levels IV and V maintenance are planned to be combined, and for any type $e$, we have $E = S^{\text{Type}}$, $G = \{4,5\}$, thus $\lambda^{e4} = 1$ and $\lambda^{e5} = 1.25$.

The concept of economies of scale, as observed in shared workshop facilities, indicates a reverse correlation between the quantity of maintenance lines at a base and the per-unit investment cost needed for maintenance capacity. This suggests that with an increase in the number of maintenance lines, the investment cost per maintenance capacity unit tends to decrease. Initial establishment of high-level maintenance shops incurs a fixed cost, which grows linearly with each added maintenance line. Table 3 shortlists 11 feasible alternative projects, along with the number of maintenance lines (positions) they possess, and provides the investment amounts for each plan referencing Xi'an base, where expansion plans are exclusive to Xi'an base.

**Table 3 Maintenance lines (positions) scale and reference investment amounts for projects**

| $p$ | CRH | | CR | | $I_7^p$ /M RMB | |
|---|---|---|---|---|---|---|
| | III | IV&V | III | IV&V | New | Expansion |
| 0 | 0 | 0 | 0 | 0 | ¥0 | ¥0 |
| 1 | 3(6) | 0 | 0 | 0 | ¥1,000 | ¥400 |
| 2 | 0 | 0 | 3(6) | 0 | ¥1,000 | ¥400 |
| 3 | 3(6) | 0 | 3(6) | 0 | ¥1,400 | ¥800 |
| 4 | 3(6) | 3(6) | 0 | 0 | ¥2,500 | ¥1,100 |
| 5 | 1.5(3) | 1.5(3) | 0 | 0 | ¥1,950 | ¥450 |
| 6 | 0 | 0 | 3(6) | 6(12) | ¥3,200 | ¥1,800 |
| 7 | 0 | 0 | 1.5(3) | 3(6) | ¥2,300 | ¥900 |
| 8 | 1.5(3) | 1.5(3) | 1.5(3) | 3(6) | ¥2,850 | ¥1,450 |
| 9 | 1.5(3) | 1.5(3) | 3(6) | 6(12) | ¥3,750 | ¥2,350 |
| 10 | 3(6) | 3(6) | 1.5(3) | 3(6) | ¥3,400 | ¥2,000 |
| 11 | 3(6) | 3(6) | 3(6) | 6(12) | ¥4,300 | ¥2,900 |

The operation costs for levels III, IV, and V maintenance tasks of a CRH series standard trainset are about 4 million RMB (approximately 560,000 USD), 13.5 million RMB (approximately 1.9 million USD), and 25 million RMB (approximately 3.5 million USD). CR series, having a higher degree of domestic production, has maintenance operation costs set at 70% of those for the CRH series, based on the reference train purchase prices. The facility investment amount and maintenance operation cost per trainset vary depending on factors such as the economic development level of the railway hub city where the base is located, the average city wage, land prices near the station, and logistics costs for necessary parts. Similarly, using Xi’an base as a benchmark, the variation ratios of investment amounts and maintenance costs for each alternative plan are shown in Table 4.

**Table 4 Variation ratios of investment amounts and unit maintenance operation costs for alternative bases**

| **Ratios** | **Urumqi** | **Hami** | **Xining** | **Lanzhou** | **Yinchuan** | **Baoji** | **Xi’an** |
|---|---|---|---|---|---|---|---|
| Facility investment | -10% | -12% | -4% | 6% | -2% | -4% | 0% |
| Cost per std trainset | 6% | 4% | 3% | -3% | -5% | -6% | 0% |

5.2 Optimal results

Based on the parameters mentioned, the model constructed in this paper is linear and was solved using Python 3.6 to invoke CPLEX 12.8.0. All calculations were performed on a computer equipped with an Intel(R) Core (TM) i5-1035G4 CPU and 8.00 GB of memory. The optimal solution obtained is: constructing a new base in Hami and expanding Xi’an base, with both selecting the investment plan $p = 8$.

The total investment for this optimal solution is 3,958 million RMB, with an annualized construction cost of approximately 359.21 million RMB and an annual total cost for high-level maintenance of EMU trains of 1,918.94 million RMB, resulting in an optimal value of 2,278.15 million RMB for the upper-level planning model.

Following the completion of the investment and construction according to this plan, trains from Urumqi and Hami depots will no longer need to travel to Xi’an for maintenance but will instead choose the newly built Hami base for high-level maintenance operations. Trains from Baoji South, Xi’an North, and Xi’an East depots will continue to use Xi’an base for maintenance tasks, while those requiring maintenance from Xining, Lanzhou West, and Yinchuan depots will be dispatched to Hami and Xi’an bases respectively. Hami base will undertake maintenance tasks for 57 standard trainsets, and Xi’an base will handle 137 standard trainsets. The specific allocation of maintenance tasks is visualized in a scatter plot shown in Figure 5.

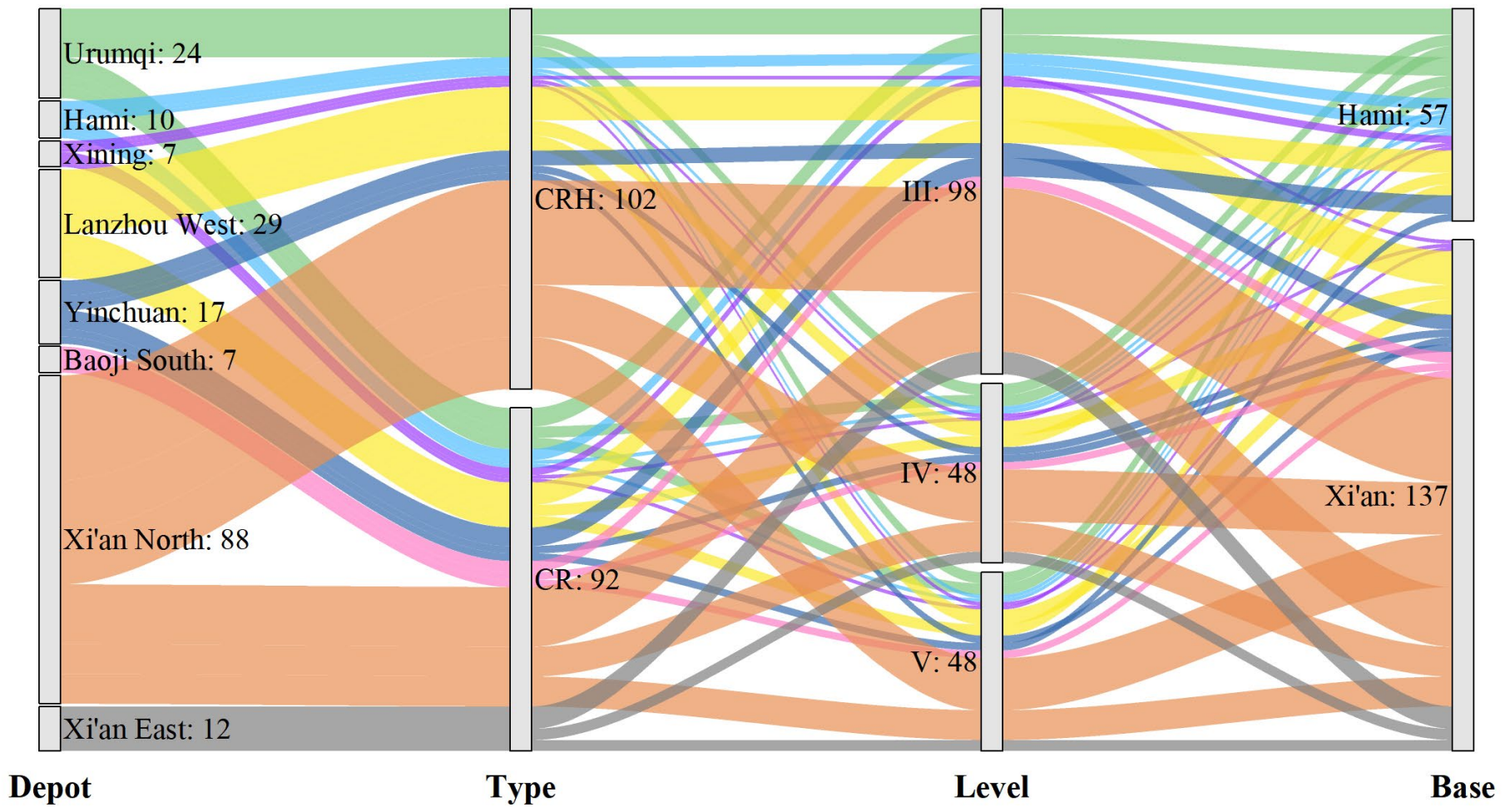

**Figure 5 Allocation of maintenance tasks**

After the investment, both maintenance bases are equipped as HMF-III&IV&V. Hami base has an annual capacity for level III maintenance of 60 standard trainsets (30 sets each for CRH and CR series) and a level IV&V maintenance capacity of 60 standard trainsets (20 sets for CRH series and 40 sets for CR series), incurring an annual maintenance operation cost of 446.59 million RMB. Xi'an base, on the other hand, can handle 90 standard trainsets for level III maintenance (60 sets for CRH series and 30 sets for CR series) and 100 standard trainsets for level IV&V maintenance (60 sets for CRH series and 40 sets for CR series), with an annual maintenance operation cost of 1,284.15 million RMB.

Figure 6 provides a comparative analysis of dispatch and maintenance operation costs for each depot, contrasting the optimized scenario of this study with the previous arrangement dependent only on Xi'an base. After implementing the optimization strategy, we observe that the new Hami base, despite its higher unit maintenance operation costs due to logistics challenges, offers a strategic advantage in terms of geographical location. This location significantly lowers the dispatch costs for nearby depots. For example, trains that previously required long-distance transport to Xi'an for maintenance can now be serviced at the closer Hami base, resulting in substantial savings on dispatch expenses. The line chart in Figure 6 clearly illustrates a notable reduction in dispatch costs for the Urumqi and Hami depots. In contrast, with the original setup relying solely on Xi'an base, all depots would have to transport their trains to Xi'an for high-level maintenance. Although unit maintenance operation cost of Xi'an might be marginally lower, the overall allocation expenses would rise substantially due to the need for long-haul transportation of trains, especially for depots situated at a considerable distance from Xi'an. Meanwhile, for Baoji South, Xi'an North, and Xi'an East depots, which chose to conduct all train maintenance at Xi'an base both before and after the optimization, their dispatch and maintenance operation costs remained unaffected.

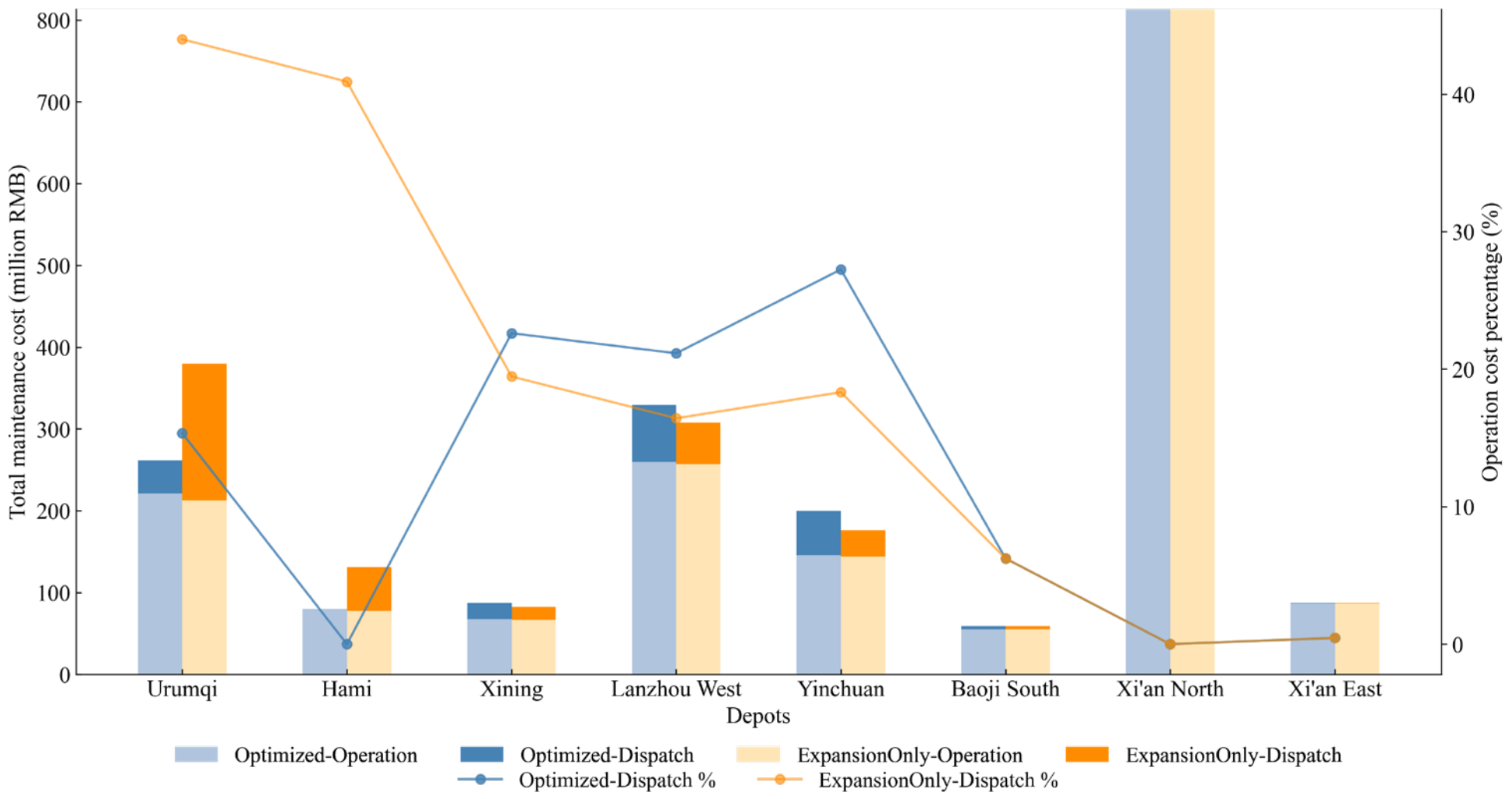

**Figure 6 Dispatch and maintenance operation costs of depots in optimized and expansion-only scenario**

This cost optimization results in a range of benefits. Economically, it makes the overall system more cost-effective. Additionally, it offers other potential advantages like reduced turnaround time for EMU trains, shorter maintenance response times to enhance safety, and enhanced operational efficiency of trains.

CSRG imposes strict policy restrictions on the number of maintenance bases, meaning in general, we are only permitted to choose the expansion plan at Xi'an base. Corresponding to Table 3, the chosen plan for Xi'an base is $p=11$ for expansion, with a facility construction cost of only 2,900 million RMB. This is significantly lower than the 3,958 million RMB required for the new Hami base project. However, while this option might appear economically advantageous when compared to others, it is not necessarily the optimal choice. The lower initial construction cost at Xi'an must be weighed against other factors, such as potential long-term operational and dispatch costs, efficiency, and capacity constraints. After considering all factors, the annual total cost of expanding only Xi'an base reached 2,300.56 million RMB, which is about 10% higher than the optimized solution we proposed, and will accumulate more expenses in long-term operation. Therefore, a comprehensive evaluation using the bi-level model we have developed is essential to determine the most beneficial approach.

Our study also finds that the annualized cost of capital investment generally accounts for only about 10%-20% of the total annual costs. Although the impact of budget constraints on the overall cost allocation might not be significant, as Melo et al. [33] mentioned, the lack of budget constraints in the opening of facilities (they also mentioned closure and relocation) can theoretically oversimplify the problem. In fact, budgetary constraints on one-time investments have a minor impact on the results of MBLAP optimization model. This phenomenon reveals a key insight: in long-term operations, the impact of annual operating costs may far exceed the initial capital investment. This suggests that in optimizing MBLAP, the focus should shift more towards enhancing operational efficiency and reducing long-term costs, rather than concentrating solely on the magnitude of initial investments. Although the budget for facility infrastructure construction is a primary concern for operators, this revelation is significant for decision-makers. It guides them to prioritize the management and control of operational costs in their resource allocation and optimization decisions.

## 5.3 Sensitivity analysis

CSRG has been advocating reforms in the maintenance processes and scheduling for locomotive vehicles. This involves setting the EMU train mileage cycle threshold and the maintenance duration. In this context, a sensitivity analysis evaluates the impact of various factors on the total costs of EMU trains. These factors include the train fleet size, the mileage cycle threshold, and the maintenance duration. The analysis uses the optimization findings from this study as a reference point. Each factor is adjusted proportionally (ranging from -20% to 20%, in increments of 5%), and the impact on the annual total cost and the proportion of construction cost within it, are monitored. Based on our prior analysis, budget constraints have a relatively small impact on annual total costs. So, when expanding the scale, we slightly loose these budget constraints to prevent situations where no viable solutions can be identified. The outcomes of this analysis are depicted in Figure 7.

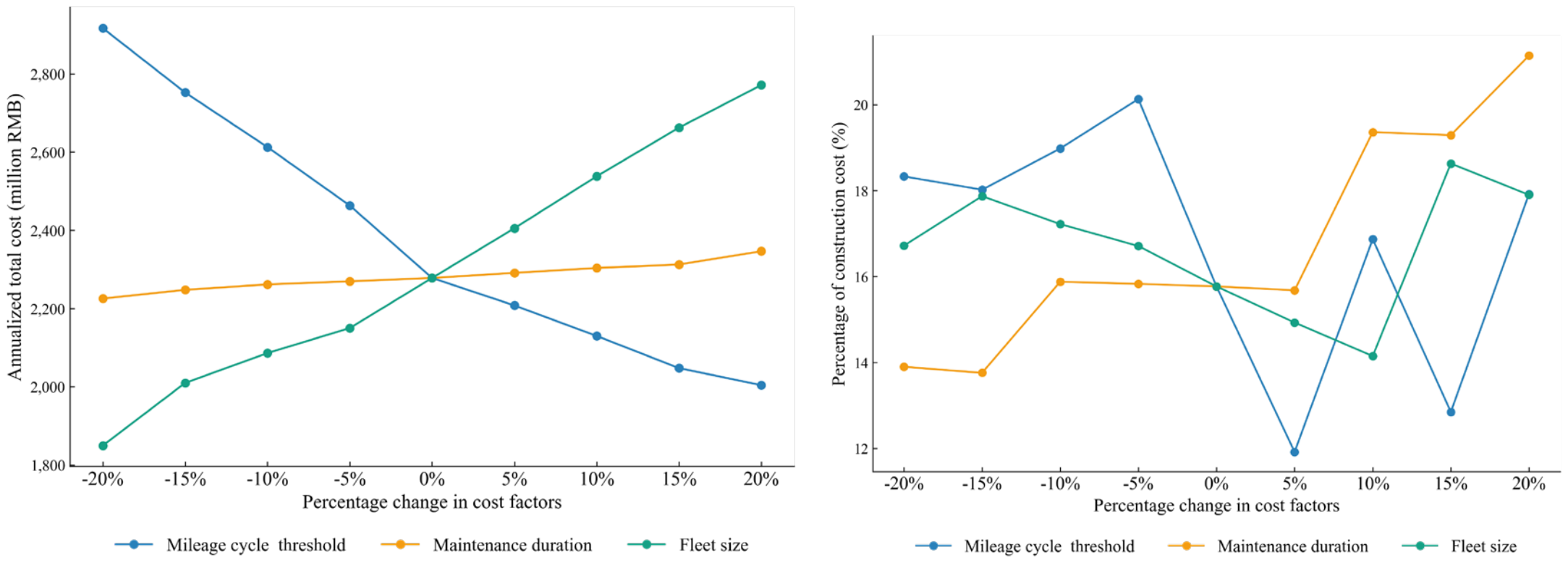

**(a) Sensitivity analysis of annualized total cost** **(b) Sensitivity analysis of percentage of construction cost**

**Figure 7 Sensitivity analysis of annualized total cost**

The analysis of the results can be summarized as follows:

1. Fleet size: The green line in the analysis graphically illustrates a continuous and sharp increase in total maintenance costs as the train fleet expands. This is a predictable outcome, as a larger number of standard trainsets naturally incurs higher annualized total costs.

2. Mileage cycle threshold: Illustrated by the blue line, there is a notable downward trend in costs as the mileage cycle threshold increases. However, the rate of decrease in costs is less pronounced compared to the impact of the train fleet size. This indicates that extending the high-level maintenance mileage cycle threshold can effectively reduce the costs associated with optimizing the EMU maintenance framework. Yet, it is crucial to balance this with the potential rise in failure rates that may accompany increased mileage limits.

3. Maintenance duration: Represented by the orange line, there is a relatively steady decline in costs when the maintenance duration is reduced. This reduction reflects the efficiency gains and benefits of shortening the period for high-level maintenance. However, the influence on the total costs is not as marked as the changes in mileage cycles. This implies that the annualized total costs are comparatively less sensitive to variations in the maintenance duration.

The overall cost of MBLAP is predominantly linked to the train fleet size. However, refining the mileage cycle threshold and the maintenance duration can also lead to cost reductions. Notably, adjustments to mileage cycles have a more pronounced effect than alterations in maintenance duration. It is crucial to possess high-level maintenance capabilities that align with the fleet size. Selecting strategic locations for maintenance bases and effectively managing the train fleet size in response to anticipated demand and budgetary constraints are essential strategies for efficient cost management.

Regarding the construction cost, its proportion of the total cost exhibits fluctuations, particularly with changes in the mileage cycle threshold, indicating different economic efficiencies under various combinations of maintenance base location choices. The proportion of construction cost gradually decreases as the fleet size increases from -20% to 10%, but then rises again with 15% and 20% increases due to choosing larger capacity construction plans (opting for plan $p=8$ to build Urumqi base and plan $p=11$ to expand Xi'an base). This might suggest that with an increase in scale, the unit construction cost is overall decreasing. Moreover, there is a discernible positive correlation between the proportion of construction cost and the extension of maintenance duration, which is an observation of particular interest. While the total cost is not markedly affected by variations in maintenance duration, the construction cost is significantly sensitive to these changes. Even though the annualized construction cost is a smaller portion of the total annual cost, this finding is significant for stakeholders such as government agencies focused on controlling construction costs, and building and engineering consulting firms. By optimizing maintenance duration settings, these entities stand to gain significant benefits.

## 6 Conclusions

This study presented an innovative approach to optimizing the location of maintenance bases for EMU trains in high-speed railway network of China. Utilizing a novel bi-level model to solve the location-allocation problem, the research effectively balanced capital investment and operational costs in strategic maintenance base planning. The model established in this paper, applied to the Northwest high-speed rail network which solved by CPLEX, provides strategic decisions for establishing a new base in Hami and expanding the existing base in Xi'an. This optimization resulted in a total investment of approximately 3,958 million RMB, with the annualized construction cost being around 359.21 million RMB and the annual high-level maintenance cos of EMU trains at about 1,918.94 million RMB. The model effectively balances the capital investment and operational costs, demonstrating its potential in guiding the construction of maintenance facilities for high-speed rail and enhancing the reliability of the network system.

It is important to acknowledge that this research primarily focused on a simplified prediction of train fleet sizes and their maintenance requirements. We did not delve into the varying maintenance needs arising from different operational methods of each train. Future studies should aim to employ scientific methodologies for a more detailed and accurate forecast of the maintenance needs of trains, which would significantly enhance the precision and reliability of maintenance planning in high-speed rail networks.

## Acknowledgments

The research was supported by the National Natural Science Foundation of China (U2268207).